\documentclass[AMA,STIX1COL
]{WileyNJD-v2}

\usepackage[utf8]{inputenc}
\usepackage{siunitx}    
\usepackage{csquotes}
\usepackage[capitalize]{cleveref}

\articletype{Article Type}%

\received{}
\revised{}
\accepted{}

\begin{document}

\title{Digital Contact Tracing Service:\\ An improved decentralized design for privacy and effectiveness }

\author[1]{Kilian Holzapfel*}
\author[1,2]{Martina Karl*}
\author[1]{Linus Lotz*}
\author[3]{Georg Carle}
\author[4]{Christian Djeffal}
\author[1]{Christian Fruck}
\author[1]{Christian Haack}
\author[5]{Dirk Heckmann}
\author[6]{Philipp H. Kindt}
\author[3]{Michael Köppl}
\author[1]{Patrick Krause}
\author[2]{Lolian Shtembari}
\author[4]{Lorenz Marx}
\author[1]{Stephan Meighen-Berger}
\author[1]{Birgit Neumair}
\author[1]{Matthias Neumair}
\author[1]{Julia Pollmann}
\author[1]{Tina Pollmann}
\author[1]{Elisa Resconi}
\author[1]{Stefan Schönert}
\author[1]{Andrea Turcati}
\author[1]{Christoph Wiesinger}
\author[1]{Giovanni Zattera}
\author[7]{Christopher Allan}
\author[7]{Esteban Barco}
\author[7]{Kai Bitterschulte}
\author[7]{Jörn Buchwald}
\author[7]{Clara Fischer}
\author[7]{Judith Gampe}
\author[7]{Martin Häcker}
\author[7]{Jasin Islami}
\author[7]{Anatol Pomplun}
\author[7]{Sebastian Preisner}
\author[7]{Nele Quast}
\author[7]{Christian Romberg}
\author[7]{Christoph Steinlehner}
\author[7]{Tjark Ziehm}

\authormark{ContacTUM consortium \textsc{et al}}

\address[1]{\orgdiv{Department of Physics}, \orgname{Technical University of Munich}, \orgaddress{\state{Garching bei München}, \country{Germany}}}

\address[2]{\orgname{Max Planck Institute for Physics}, \orgaddress{\state{Munich}, \country{Germany}}}

\address[3]{\orgdiv{Department of Informatics}, \orgname{Technical University of Munich}, \orgaddress{\state{Garching bei München}, \country{Germany}}}

\address[4]{\orgdiv{TUM School of Governance, Munich Center for Technology in Society}, \orgname{Technical University of Munich}, \orgaddress{\state{Munich}, \country{Germany}}}

\address[5]{\orgdiv{TUM School of Governance | Department of Informatics, Chair of Law and Security in Digital Transformation}, \orgname{Technical University of Munich}, \orgaddress{\state{Munich}, \country{Germany}}}

\address[6]{\orgdiv{Department of Electrical and Computer Engineering, Chair of Real-Time Computer Systems}, \orgname{Technical University of Munich}, \orgaddress{\state{Munich}, \country{Germany}}}

\address[7]{\orgname{ito-app}, \orgaddress{\state{Ober-Olm}, \country{Germany}}}

\corres{*Kilian Holzapfel, Martina Karl, Linus Lotz, Department of Physics, Technical University of Munich, James-Frank-Str. 1, D-85748 Garching bei München, Germany. \\ \email{Kilian.Holzapfel@tum.de}, \\ \email{Martina.Karl@tum.de}, \\ \email{Linus.Lotz@tum.de}}


\abstract[Summary]{We propose a decentralized digital contact tracing service that preserves the users' privacy by design while complying to the highest security standards. Our approach is based on Bluetooth and measures actual encounters of people, the contact time period, and estimates the proximity of the contact. We trace the users' contacts and the possible spread of infectious diseases while preventing location tracking of users, protecting their data and identity. We verify and improve the impact of tracking based on epidemiological models. We compare a centralized and decentralized approach on a legal perspective and find a decentralized approach preferable considering proportionality and data minimization.}




\maketitle

\section{Introduction}\label{sec:introduction}
The current COVID-19 pandemic poses challenges to our society on a scale unheard of in recent times. While the direct consequences of the pandemic are felt in healthcare and medical sectors, the quarantining and isolation measures required to slow the outbreak have a major impact on the psychological and economic welfare of people. One measure that has been applied in a digital and analogue manner is the tracing of contacts. State authorities resorted to this measure in order to find out about new infectious persons and prevent further spreading of the disease by quarantining them. Many of the analogue solutions are problematic because state authorities might not have the capacity to question and pursue contacts and when they do, they touch upon the privacy of citizens. A first wave of governmental apps have been criticized on the basis that they could be used for surveilling citizens. As a reaction, more privacy friendly approaches have been proposed. The experiences with existing apps have shown that potential loopholes are exploited even by citizens in order to identify infected persons and e.g. shame them on social media \cite{SouthCoreaNature2020} \cite{SouthCoreaGuardian2020}. Hence, the issue arose whether there is an effective digital solution that also takes into account ethical, legal, and societal concerns. Therefore, our concept puts forward ideas to improve the decentralised concept. 

Our current proposal aims at fighting infectious diseases in an effective manner while safeguarding and realizing the citizens’ rights, freedoms and legitimate interests. We focus on privacy and IT-security concerns. In the spirit of a privacy by design solution, we incorporated legal principles and requirements into the very design of our solution. While our first point of reference was the European Union’s General Data Protection Regulation (GDPR) 2016/679, there exist similar and equivalent principles and requirements in many legal orders, including the Council of Europe’s Convention 108. These include: 
\begin{itemize}
\item lawfulness, fairness and transparency (Art. 5 Sec. 1 Subsec a. GDPR)
\item purpose limitation: data shall be \enquote{collected for specified, explicit and legitimate purposes and not further processed in a manner that is incompatible with those purposes} (Art. 5 Sec. 1 Subsec b. GDPR)
\item data minimization: data shall be \enquote{adequate, relevant and limited to what is necessary in relation to the purposes for which they are processed} (Art. 5 Sec. 1 Subsec c. GDPR)
\item accuracy: data shall be \enquote{accurate and, where necessary, kept up to date; every reasonable step must be taken to ensure that personal data that are inaccurate, having regard to the purposes for which they are processed, are erased or rectified without delay} (Art. 5 Sec. 1 Subsec d. GDPR)
\item storage limitation: data shall be “kept in a form which permits identification of data subjects for no longer than is necessary for the purposes for which the personal data are processed” (Art. 5 Sec. 1 Subsec e. GDPR)
\item integrity and confidentiality: \enquote{processed in a manner that ensures appropriate security of the personal data, including protection against unauthorised or unlawful processing and against accidental loss, destruction or damage, using appropriate technical or organisational measures} (Art. 5 Sec. 1 Subsec f. GDPR)
\end{itemize}
While the concept laid out here addresses many principles and requirements, its actual implementation will contain further technical and organizational measures in order to mitigate risks and further issues. This is particularly true for data subjects rights like the right to rectification and the right to erasure. A further development of the application will also have to look into other issues like inclusion, fairness, transparency and effective governance of the application. \\

We present a secure solution for a digital contact tracing service (DCTS) that protects the users' privacy, identity and personal data from attackers.
Encounters, their proximity, and duration are required in order to properly track contacts of people and infection chains. We propose the use of Bluetooth
, a short range wireless communication protocol, as a means to measure these quantities. Bluetooth detects only real encounters and works indoors as well as outdoors (e.g. underground in subways or in buildings), where location (e.g. by GPS) and mobile network data is not reliable anymore. 
Bluetooth is a technology standard available on every mobile phone and thus provides the ideal global instrument to register encounters on local devices.
We present the general concept of DCTS in \cref{sec:DCTS_concept}, the technical details and implementation aspects in \cref{sec:technical_details}, and consider possible attack scenarios in \cref{sec:attacks}. 
We cover related work in \cref{sec:related_work}, and  provide a legal perspective in \cref{sec:legal_aspects} and conclude in \cref{sec:conclusion}.

\section{Digital contact tracing via a Bluetooth app -- Concept}\label{sec:DCTS_concept}
In this section, we outline the general concept of DCTS. Details for the technical implementation are covered in \cref{sec:technical_details}.

We propose a mobile Bluetooth application (app) to introduce DCTS in the society. Our proposed app permits the registration of relative encounters while preserving the privacy of its users by design. The concept is based on the following principle: Each mobile phone equipped with the DCTS app advertises temporary contact numbers (TCNs) to other phones. At the same time, it records and stores the TCNs advertised by other phones. Phones continuously advertise their random TCNs and store the observed random TCNs of neighboring devices, while users can simply follow their daily routine activities (such as office, school, theater, etc.). In case users are infected, they can agree to an upload of their advertised TCNs to a server after approval from medical authorities. Every app user continuously checks the server and gets information on the TCNs related to people tested positive for the virus. A matching operation done on the user's device reveals them to the user only if a potentially infectious contact has happened. In this way, each user is informed about potential infectious contact without revealing so to another party. The identity of the infectious person and their social graph remains protected. 

\subsection{Generation and advertising of random TCNs}
The user installs the DCTS app. The app activates Bluetooth and generates a key, which it uses to generate a random TCN. The phone then proceeds to advertise the random TCN via Bluetooth, such that other devices in the vicinity of the user can see the TCN. This TCN is updated after a certain time in order to minimize re-identification of the user. The app stores the advertised TCNs for a period of two or three weeks, depending on what is sensible for the infectious period of the virus. The key for TCN generation is updated every day and is stored on the phone. This ensures that key compromising can not deanonymize users' past movements.

\subsection{Scanning and storing TCNs of neighboring phones}
\begin{figure}
    \centering
     \includegraphics[width=\textwidth,page=1]{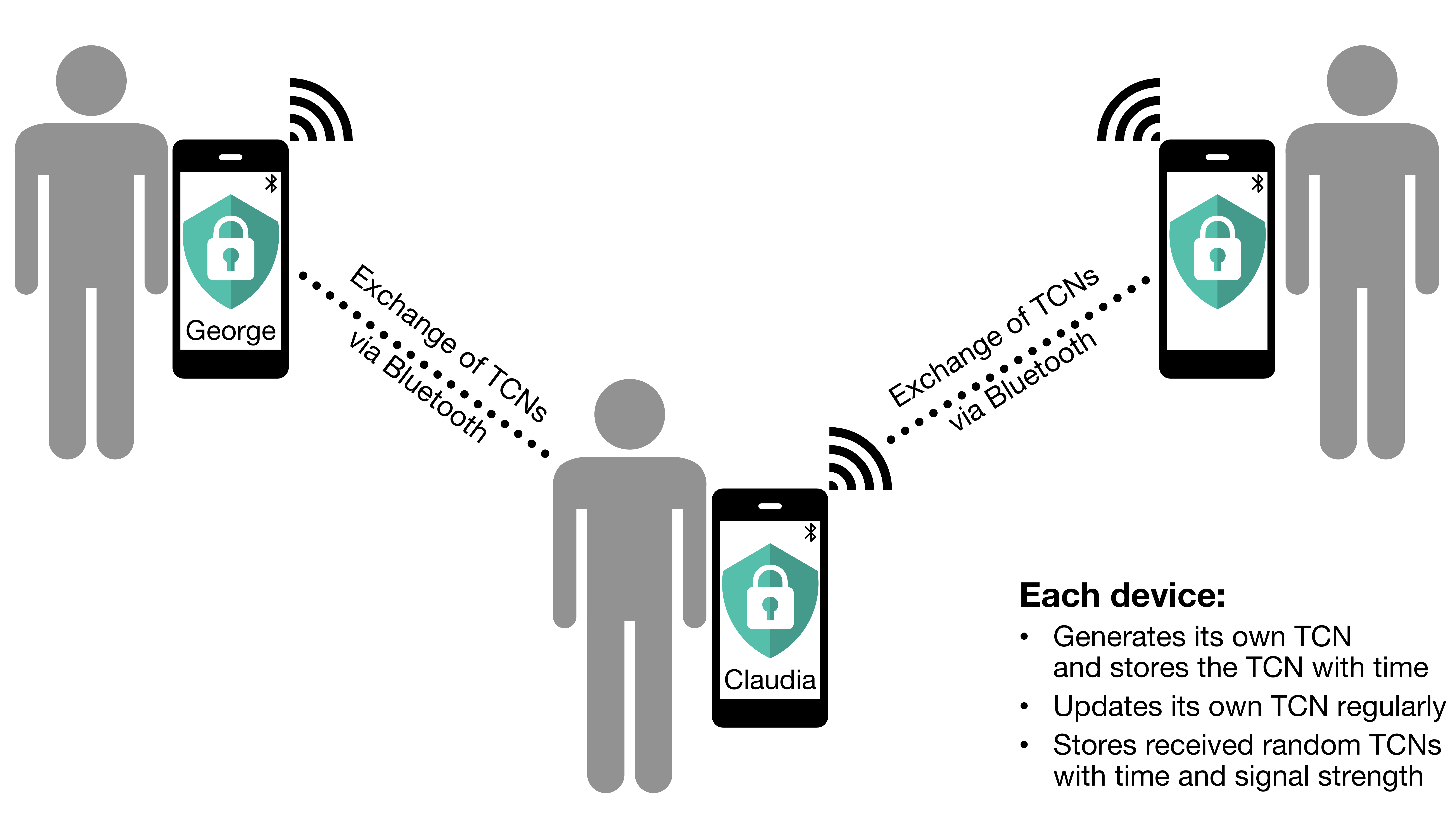}
    \caption{Phones in each others' vicinity exchange their random TCNs via Bluetooth. All TCNs are updated after a certain time period.}
    \label{fig:id_exchange}
\end{figure}

In parallel to the advertising operation, the Bluetooth activated on the device continuously scans for other devices in its vicinity. When neighboring devices are detected, the app stores the observed TCNs, the time, and signal strength on the phone. The period of exposure can be calculated using the saved timestamps and the proximity can be evaluated based on the received signal strength. We only store encountered TCNs for a time period of two or three weeks. We show a sketch of the TCN exchange between different phones in \cref{fig:id_exchange}.

\subsection{Uploading TCNs to the server}
\begin{figure}
    \centering
    \includegraphics[width=\textwidth,page=2]{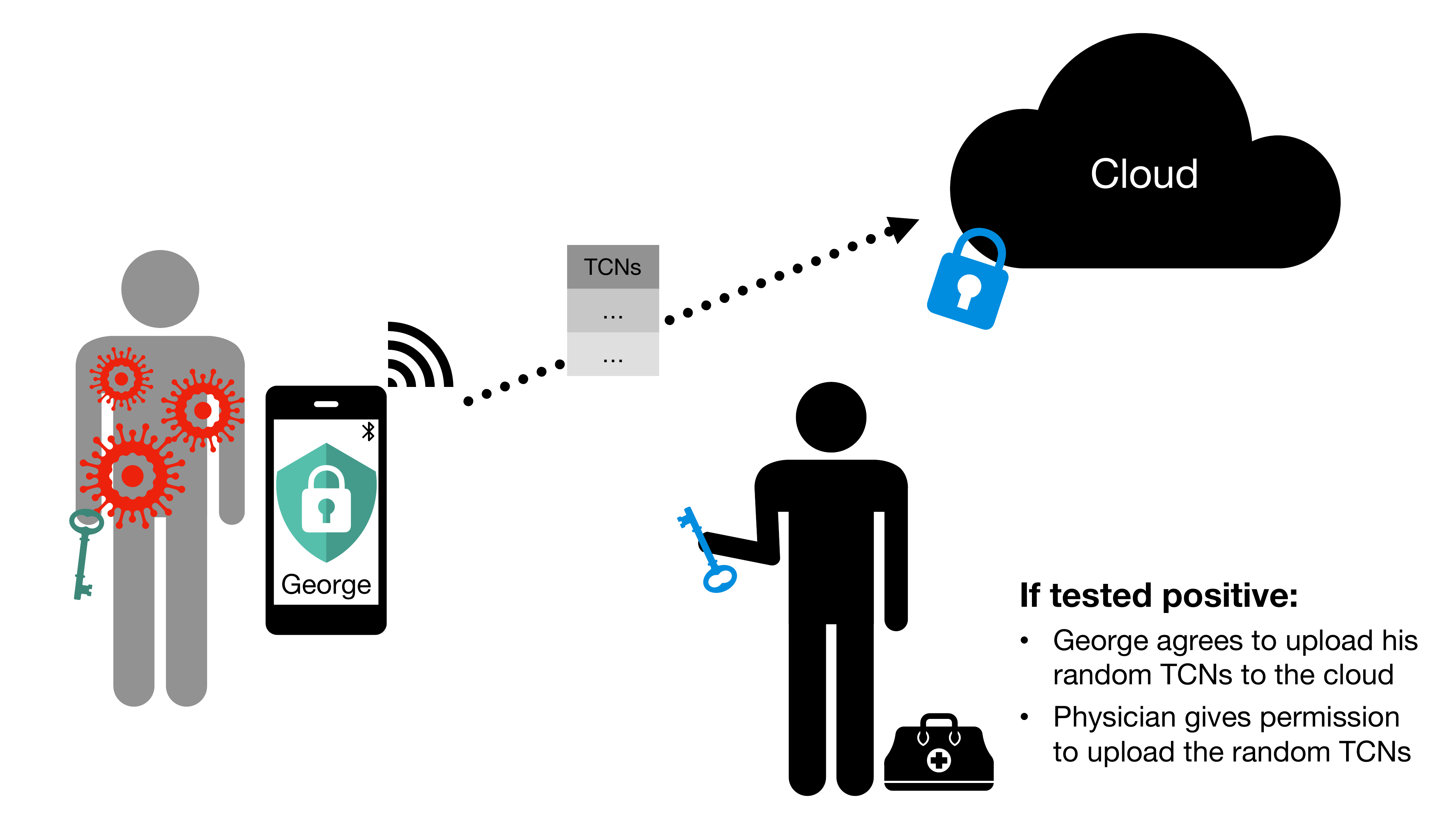}
    \caption{Scenario 1: When a user receives a positive test result, they get the permission to upload their random TCNs from the past two or three weeks to the server.}
    \label{fig:id_upload}
\end{figure}

If a user is tested positive for the virus, the patient is encouraged by medical authorities to provide the TCNs advertised over a period of two or three weeks. The patient is informed about how their identity is protected, including the description of the risk of identification via possible attacks (see \cref{sec:id_infected_person}).
If the patient agrees, their advertised TCNs are uploaded to a server where they are verified and encrypted before being made available. 
 \begin{itemize}
     \item \textbf{Scenario 1: Patient uploads advertised TCNs and keys} \\
The patient gets the permission by a medical authority to upload the generated TCNs, and keys to a server. This permission can be granted in various ways (see \cref{sec:technical_details}). The patient then proceeds to upload the keys. The server regenerates and verifies the patient's TCNs with the provided keys. This verification prevents impersonation of other users. The server deletes the keys after the verification. This scenario is shown in \cref{fig:id_upload}.\\
     \item \textbf{Scenario 2: Doctor uploads advertised TCNs} \\
The patient provides the keys used for TCN generation to the medical personnel, e.g. by showing a QR code with the keys to an authorized person. The medical personnel then verifies the TCNs by regenerating them and uploads the TCNs to the server following an authentication procedure (e.g. username and password of medical authority). The medical personnel deletes the keys after the upload.
 \end{itemize}

Scenario 2 ensures the user's anonymity towards the server and does not require to use mobile data or WiFi. The app generates a new key for future TCN generation after the upload.

\subsection{Collecting TCNs on the server}
The server collects newly uploaded TCNs for a predefined period of time, e.g. one hour, and shuffles their order to avoid the association of several TCNs to a single user. Then, the server stores the shuffled batch of TCNs to its main database. This enables users' apps to check whether they were in contact with the patient. TCNs are only stored for two/three weeks and are then automatically deleted.

\subsection{Discovering potentially infectious contacts and slowing down the pandemic}
\begin{figure}
    \centering
    \includegraphics[width=\textwidth,page=3]{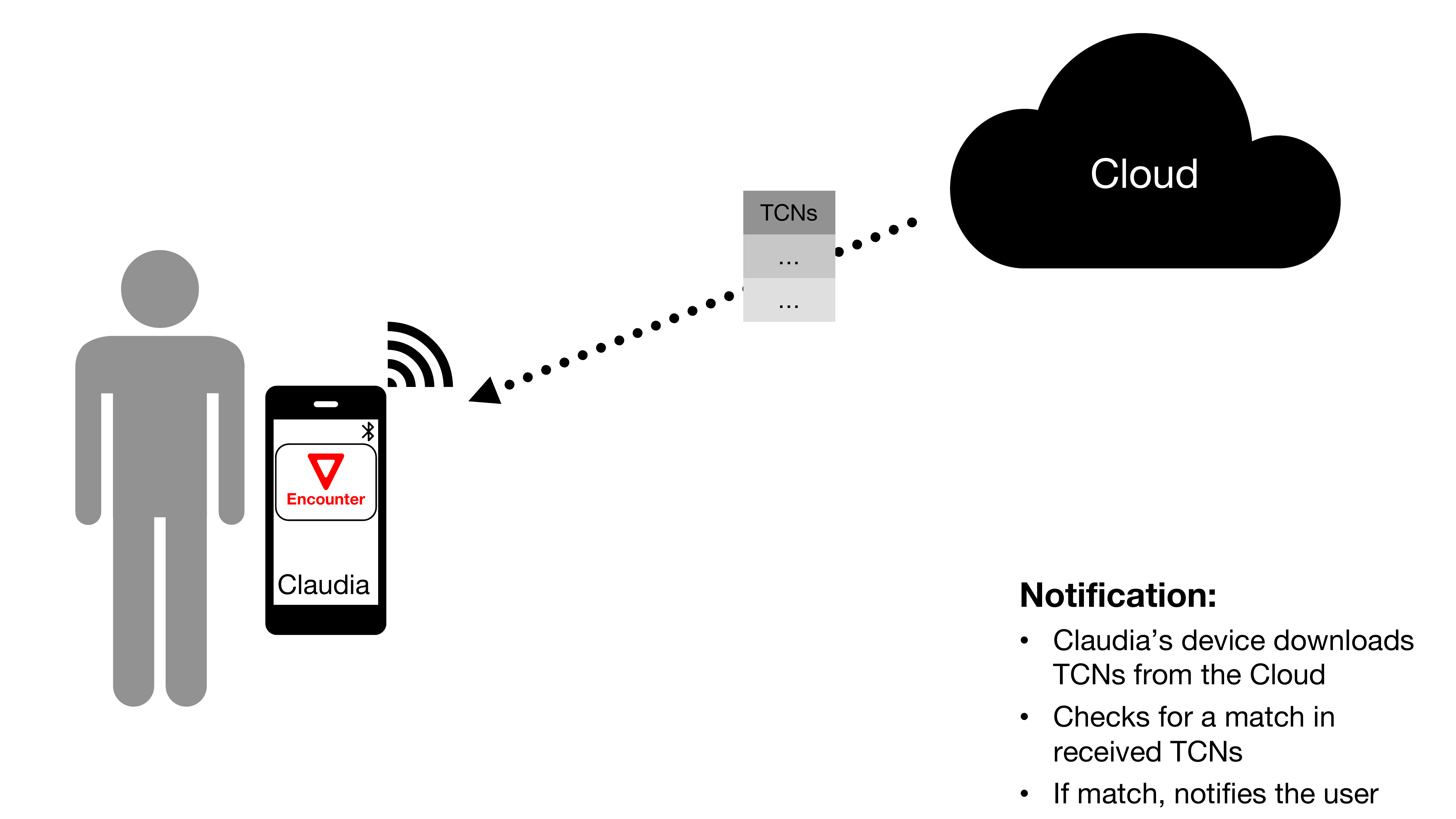}
    \caption{Every device can check its recorded TCNs against the reported TCNs on the server. If a device finds a match, it notifies the user.}
    \label{fig:notification}
\end{figure}

Users in their daily life have the app working on their devices. In parallel to the continued advertisement of TCNs, the app checks regularly, e.g., once per hour, for new TCNs on the server. If new TCNs are present, the app retrieves information about them from the server. The patients' TCNs are then matched against the encountered TCNs registered on the device of the user during a period of two/three weeks. This matching is done on the user's phone.
If the app detects a match, the user receives a notification that a potentially infectious encounter has been detected (\cref{fig:notification}).

This notification includes recommended actions, such as self-quarantining and calling a number or visiting a website with contact details for medical authorities. The notified user can also be asked to proceed and provide his/her TCNs to allow a recursive tracing. We recommend consulting psychologists about the exact wording and information of this notification in order to achieve the desired effect. We present two approaches on how to detect encounters in the following.

\subsubsection{Direct download of TCNs}\label{sec:direct_download}
In order to check whether the user has been in contact with an infected person, they download all unchecked TCNs stored on the server and check for matches within their own list of observed TCNs. When encountering matches, the app can perform a risk assessment based on exposure time period and proximity. The risk assessment can be included in the notification. This approach is vulnerable to an attack described in \cref{sec:id_infected_person}. A possible solution for this vulnerability is described in the following section.

\subsubsection{Checking for overlap of scanned TCNs and infected TCNs}
The previous scenario is vulnerable to the attack scenario described in \cref{sec:id_infected_person}. An attacker can put a device with the app together with a video camera at a public place, record the broadcasted TCNs and can later check for infected TCNs on the server. If the attacker has recorded an infected person, they can potentially assign infected TCNs to people on the camera. In order to avoid this attack, we can check for the number of TCNs that are both in our set of encountered TCNs and in the set of infected TCNs. An algorithm determining the private set intersection cardinality with low communication cost (see for example  \cite{private_set_2018}) is a valuable strategy in order to discover the exposure to infectious contacts without risking the identification of the patient (details are described in \cref{sec:private_set_intersection}). 

If intersection cardinality is used, the risk assessment can still be done by e.g. introducing three categories: high exposure, medium exposure, and low exposure. The encountered TCNs are sorted into these categories, depending on the contact period and the proximity. Then the number of overlapping TCNs in each category can be checked, which provides a measure for user exposure.

\subsection{Second order contact tracing: tracing contacts of traced contacts}
Several studies have confirmed that COVTID-19 is infectious before people develop symptoms \cite{incubationPeriod, Rothe2020}. This leads to a spread of the virus before people get diagnosed and can isolate. 

For example, person A gets infected. Before developing symptoms, person A infects person B. Person A still shows no symptoms and B continues living normally and infects person C. Now A develops symptoms. After A got the diagnosis, person B is notified, but person C is still oblivious. Thus, it is possible for the virus to spread much faster than direct contacts can be traced. In \cite{contactumodel}, it is shown that infection chains can only be stopped if indirect contacts are traced as well.

We enable person B to upload their TCNs to the server as soon as B gets the notification of having been exposed. If B uploads their TCNs, person C gets a notification before infecting anybody else. For more details see \cref{sec:2ndorder}.

\section{Technical details}\label{sec:technical_details}
We explain the details of implementation and protocols in this section. A more general overview of the concept is provided in \cref{sec:DCTS_concept}.

\subsection{Link Layer Operation}
\label{sec:bluetooth}
Our proposed approach uses Bluetooth Low Energy (BLE) for detecting devices in range.
The procedure using which two wireless devices establish a first contact in a wireless network is called neighbor discovery. In BLE, devices periodically broadcast packets with an interval $T_a$ for neighbor discovery. For reducing the probability that multiple consecutive packets of different devices are sent at the same point in time and hence collide, a random delay between $0$ and $\SI{10}{ms}$ is added to each instance of $T_a$. In addition, devices listen to the channel for a time window of length $d_s$ every $T_s$ time-units. A device has successfully discovered another one, once a beacon from the opposite device coincides with one of its reception windows.

We have estimated the performance when two smartphones discover each other (cf. \cite{kindt:20} for details). The choices of  $T_a$, $T_s$ and $d_s$ supported by the Android operating system are not officially documented. We have therefore looked them up in the source code of the Android operating system. Due to scheduling conflicts, the values actually used could differ during runtime. We nevertheless found that for certain configurations, the latency measured from the point in time at which two devices come into range until discovery is successful is below $\SI{5}{s}$ during normal operation, i.e., when no scheduling conflicts occur. Such latencies are practical for contact tracing. We also found that continuous contact tracing has no significant impact on the smartphone battery runtime. We expect that the battery is drained by no more than $\SI{5}{\percent}$ by contact tracing, while the energy demand is even significantly below that in most cases. Finally, we investigated the behavior in crowded situations, where a large number of devices are in range of reception. Here, the packets from multiple devices could potentially collide. We found that even in situations with $100$ devices being close to each other, the probability that all devices discover each other successfully within $\SI{10}{s}$ is close to $\SI{100}{\percent}$.

In our approach, we chose the most beneficial parameters for BLE based on this evaluation. We thereby ensure that contact tracing is carried out with the highest possible reliability and the lowest possible energy consumption.

Distance estimation is done by evaluating the received signal strength indicator (RSSI) provided for each received packet. This estimation is known to be error-prone. In our approach, we eliminate as many sources of error as possible, while classifying a contact as significant by jointly considering the RSSI and contact duration. This reduces the rate of false positives and negatives.

\subsection{Generating TCNs and advertising}
Our approach is similar to the contact tracing suggested by Apple and Google\cite{appleGoogleTracing2020} in this aspect. We suggest modifying their approach by using a completely random daily key ($rdKey$) everyday to ensure forward secrecy. In the original approach, a leak of the private key allows an attacker to reproduce all past and future daily tracing keys.

From \cite{appleGoogleTracing2020} the pseudo random TCN ($prTCN$) generation looks as follows:
\begin{equation} \label{eq:pseudoRandomID}
    prTCN \leftarrow \text{Truncate}(\text{HMAC}(rdKey, \text{UTF8}(\text{"CT-RPI"}) || TIN)),16),
\end{equation}
where, $TIN$ is the time interval number, the n-th 10 minute of the day (e.g. 0:22 would be in the second time interval, thus $TIN=2$). This time interval number prevents rebroadcasting the TCNs of other users in other time intervals.

 Bluetooth also advertises the MAC-address of the device. According to our observations, this address changes after a certain time and is also changed when Bluetooth is activated. We were able to observe this behaviour in our tests on Android 8 and 9 as well as iOS 12. We stop and restart advertising immediately when updating the TCN, such that the advertised MAC-addresses change at the same time as the TCNs. This prevents any malicious association of a MAC-address to several TCNs. These changing TCNs keep the user anonymous and complicate tracking.

The DCTS app stores the keys and the generated TCNs in a database (e.g., SQLite encrypted with SQLCipher \cite{SQLCipher}) that is physically present on the device.

The maximal length of advertising data for Bluetooth is 31 byte (for Bluetooth 4.x). The advertisement includes a service universally unique identifier (UUID) of 16 bit. This UUID identifies the advertisement as a DCTS advertisement to other devices. The pseudo random TCNs are then advertised as additional data with a size of 16 bytes. 

Another method for random TCN generation is presented in \cite{DP3Twhitepaper}. Their described technique is similar and we are currently evaluating which approach is the most secure and privacy preserving and offers the most protection to the user. 

\subsection{Scanning and storing of encountered TCNs}
The app registers only DCTS-advertisements of devices in the vicinity, using a filter for the service UUID in the Bluetooth scans. Simultaneously, the app advertises the user's TCNs with a size of 16 byte. For Android, advertised TCNs can be read out by the scan-callback and saved into the SQLite database. When a pseudo random TCN is seen, the app calculates a contactEventTCN ($ceTCN$):
\begin{equation} \label{eq:contactId}
    ceTCN \leftarrow \text{SHA256}(receivedTCN || \text{ISO-Date}(today) || TIN),
\end{equation}
where $TIN$ is the time interval number of the contact time. 

Additionally, the app saves the contact time and the received signal strength indicator (RSSI) of the advertisement. The proximity can be estimated using the RSSI. The contact time period can be calculated if a TCN is registered several times. Both time period and proximity can then be combined to a degree of exposure to the virus.

We note here that RSSI is a relative quantity and can differ for different chips. A possibility to calibrate RSSI could be to evaluate the range of RSSI within the first days of taking data. Within these days the user most probably has had close and distant encounters with people. This reveals an estimate of the highest and lowest range of RSSI. We could then estimate the proximity with RSSI calibrated on its maximal and minimal value. However, RSSI also depends on many factors, such as for example the orientation of the devices' antennas, whether the line-of-sight is obstructed, potentially humidity, and the channel on which a packet is sent. We will rule out these errors whenever possible. The remaining error then impacts the required contact time. If e.g. the estimated distance is lower than the actual distance, the devices would need to be longer in their vicinity in order to count as relevant contact.


\subsection{Uploading of pseudoRandomTCNs}
Only medical personnel have access or can grant access to the server to perform the upload. This minimises the misuse of reported TCNs of patients who are tested positive for the virus. The app offers an interface allowing medical personnel to either upload data to the server or to grant access to the server.

In order to ensure correct use of the app, a short instruction or training for the use of the app needs to be provided. Also, we need to identify medical personnel in order to provide them with credentials for the server upload. This can be done by either contacting test centers directly or by getting the relevant contacts from the health office, to which the infected individuals are reported to. Doctors and institutions authorized to report people as infected can then be contacted and provided with the credentials and the instruction for the use of this app. 

\subsubsection{User uploads data}
Patients are asked to upload their pseudo random TCNs after consultation with the medical personnel. The permission is provided via a token or an access code or a TAN at the doctor's office or the test center.  A QR code or a TAN are provided by the doctor to the patient. Alternatively,  the access code or TAN is provided via letter together with the test result (for recursive tracing). The code or TAN is then only valid for a single use for a restricted period of time. Inserting the TAN or scanning the QR code  triggers the upload of the keys used for generating the random TCNs for the past two/three weeks. An IP anonymization protocol such as TOR \cite{dingledine2004tor} can be used to avoid revealing the patient's IP-address to the server during the upload of the TCNs. Once the keys are on the server, the server then verifies the TCNs by regenerating them with the keys for each day and all possible time interval numbers. The server then calculates for each pseudoRandomTCN the corresponding contactEventTCN (see \cref{eq:contactId}) with the respective time interval number and date. After TCN verification, the server deletes the keys. This verification ensures that patients can not simply upload observed TCNs of other people to mark them as infected. 

If the user does not have access to WiFi, the upload of the TCNs can be done in the test center or doctor's office or over mobile data. 

\subsubsection{Medical personnel uploads data}
The user hands their keys to the doctor providing a QR code. The medical personnel receives the keys via the DCTS app and regenerates the user's pseudoRandomTCNs with the fixed time interval numbers. The pseudoRandomTCNs are then hashed together with their respective time interval numbers and date (as shown in \cref{eq:contactId}) to generate the contactEventTCNs. The medical authority then proceed to upload the user's TCNs to the server using their credentials to access the server.

\subsection{Server}
The server collects all verified TCNs for a short period of time, for example one hour. The collected TCNs get sorted or shuffled and saved into a database (for example as SQLite database encrypted using SQLCipher \cite{SQLCipher}). For the private set intersection cardinality protocol, the server computes  $BF(Enc_{pk_S}(TCN_S))$ for all shuffled TCNs (see \cref{sec:private_set_intersection}). The server allows the DCTS users to download the database (or the bloom filter) of the encrypted TCNs. 

These hourly releases assure that a set of TCNs cannot be linked to one person, because the TCNs of several people are combined and their order is changed by either shuffling or sorting. Sorting the TCNs makes sense if the user directly downloads the TCNs, such that they can do a binary search for their contactEventTCNs. The user can then do hourly queries to get timely notifications in case of a contact. If the user queries the server more sporadically, they get the newly added data since their last query. 

\subsection{Private set intersection cardinality}\label{sec:private_set_intersection}


To prevent attacks that might identify the patients, the DCTS app can use private set intersection cardinality to determine the number of infectious contacts. Several possible methods to determine the private set intersection cardinality exist, for example \cite{PrivateContactDiscovery} or the protocol described in \cite{private_set_2018}. We have not yet decided which specific protocol to use, but we present the working principle of the latter protocol.

The server $S$ has a set of infected TCNs ($TCN_S$), and the user $U$ has a set of encountered contactEventTCNs ($TCN_U$). Both user and server have locally a set of public and secret keys (user: $pk_U, sk_U$; server: $pk_S, sk_S$). The user now shuffles their $TCN_U$ and encrypts them ($Enc_{pk_U}(TCN_U)$). They then send $Enc_{pk_U}(TCN_U)$ to the server. The server also shuffles and encrypts $Enc_{pk_U}(TCN_U)$, such that the server sends $Enc_{pk_S}(Enc_{pk_U}(TCN_U))$ back to the user. We use a commutative encryption scheme, such that $Enc_{pk_A}(Enc_{pk_B}(data)) = Enc_{pk_B}(Enc_{pk_A}(data))$. The user now decrypts $Enc_{pk_S}(Enc_{pk_U}(TCN_U))$ and gets $Enc_{pk_S}(TCN_U)$. A commutative encryption scheme is for example Pohlig-Hellmann \cite{PohligHellmann} or SRA \cite{SRA1981}.

\begin{figure}
    \centering
    \includegraphics[width=\textwidth]{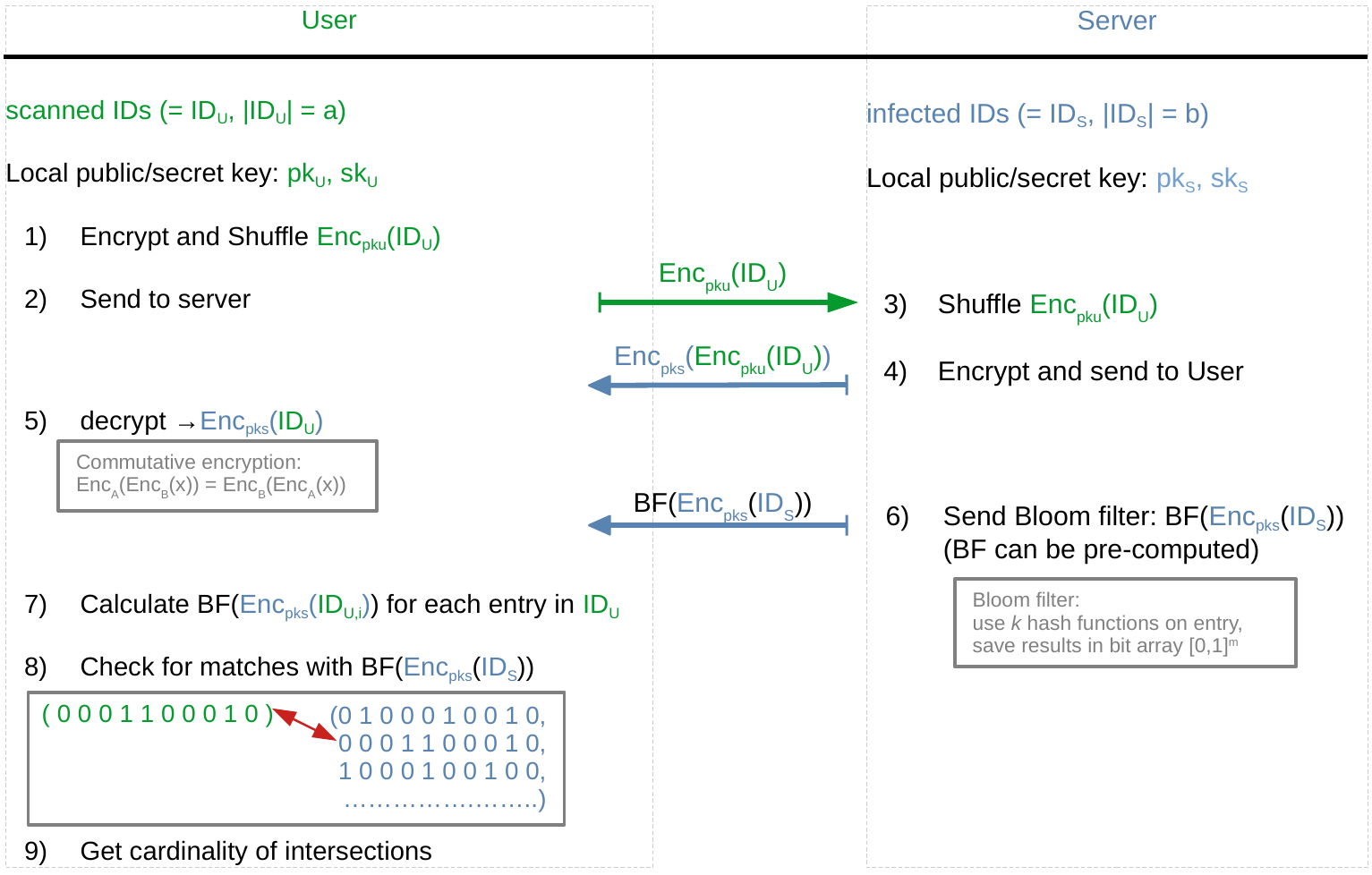}
    \caption{Scheme of the private set intersection cardinality algorithm described in \cite{private_set_2018}. With this protocol users determine how many of their observed TCNs have been marked as infected on the server. The matching happens on the users' phones, the server can not infer the number of intersections.}
    \label{fig:private_set_scheme}
\end{figure}

On the server side, the server encrypts its $TCN_S$ and applies a bloom filter ($BF$) on $Enc_{pk_S}(TCN_S)$. This step can be precomputed for all uploaded TCNs. The server then sends $BF(Enc_{pk_S}(TCN_S))$ to the user. The user also applies a bloom filter on $Enc_{pk_S}(TCN_U)$ for each of their encountered TCNs. Eventually, the user then checks whether $BF(Enc_{pk_S}(TCN_U))$ occurs in $BF(Enc_{pk_S}(TCN_S))$ for each of their encountered TCNs. The steps of the protocol are displayed in \cref{fig:private_set_scheme}.

{}

This reveals to the app how many observed TCNs are in the set of patients' TCNs saved on the server. This information is only obtained on the user's device itself, no such information is leaked to the server. The user does not know which contactEventTCN belongs to an infected person and thus we ensure the patient's anonymity. This of course makes it impossible to assign the remaining stored data, such as exposure time period or proximity, to a single direct contact. However, private set intersection cardinality does not prevent the possibility to provide a risk assessment determined on the time period and signal strength of the encounter to the user. 

In order to provide such a service, the app can check for an encounter first. Then, the contactEventTCNs are divided into sub categories, for example high exposure, medium exposure, and low exposure. The set intersections can be checked for each category and allow for a risk assessment of the user. It is sufficient to receive $Enc_{pk_S}(Enc_{pk_U}(TCN_U))$ from the server and do the set intersection with the previously received $BF(Enc_{pk_S}(TCN_S))$ for this second query step. If there was no encounter, random TCNs are picked from the observed TCNs to simulate the second query step to the server, such that the server does not know whether the users have been in actual contact with an infected person or not.
For preventing the server from recognizing the same TCNs, a new  key-pair needs to be used every time the mobile device queries the server.

The server requires a minimum number of TCNs to be queried in order to prevent the user querying single TCNs and potentially identifying infected users. 
If users only have one received contactEventTCN, the list of encountered TCNs is padded with randomly generated TCNs, such that they can still query the server. This increases the chance of a false positive result. 
We introduce a limit on the query rate, such that the app can only send a limited number of requests to the server at a time in order to prevent brute force attacks. Then users have to wait a certain period of time until they can query the server anew.

The server's public key $pk_S$ can additionally change for each combined hourly uploaded patient TCN set. This complicates a brute force attack further, because the attacker needs a different bloom filter of their encountered TCNs for each hourly dataset. 

Using this approach to private set intersection cardinality might reveal the number of encounterered TCNs to the server. This is not the case if we allow the user to directly download all TCNs of infected people and check for matches on their phone. Using the latter method, the server gets no information whatsover about non-infected app users (except maybe their IP-address), however the user could identify infected people using an attack as described in \cref{sec:id_infected_person}.

As an update to this protocol the bloom filter can be replaced with a cuckoo filter, which has the benefit of having lower error probabilities for the same size. This is also what is used in \cite{PrivateContactDiscovery}. 




\subsection{Second-order contact tracing: tracing  contacts of traced contacts}\label{sec:2ndorder}

The ContacTUM group is evaluating the efficiency of DCTS together with different intervention strategies. The results are being crosschecked using both deterministic and Monte Carlo based model approaches \cite{contactumodel}.

The modeling substantiates the following prerequisites of a DCTS. First of all, the DCTS needs a broad acceptance among the population of more than 70\% in order to have an impact to control an outbreak. We believe this can only be achieved by a decentralized, secure, and privacy preserving design where the users own their data. Our goal is to contribute to slow down and eventually to stop the spread of the pandemic using the means of contact tracing complying with privacy laws, IT-security standards and the protection of human rights.

Second of all, tracing of people who were in direct contact with a confirmed infectious person might not be sufficient. SARS-CoV-2 can be transmitted pre-symptomatic \cite{Mizumoto2020}\cite{Rothe2020} and a significant fraction of cases is asymptomatic  \cite{Mizumoto2020}\cite{Corman2020}\cite{Nishiura2020}.  To illustrate the pre-symptomatic transmission we consider the following case: person A has been infected with SARS-CoV-2. A can spread the virus to person B already up to 2-3 days before symptom onset. B is now a first order contact of A. With a non-negligible likelihood, person B can expose person C to the virus (which had no contact to Person A) before person A develops symptoms. C is thus a first order contact of B and a second order contact of A. Once person A develops symptoms and is positively tested, person B is notified after having already spread the virus to person C. This emphasizes the need for digital contact tracing and the necessity to not only notify Person B promptly, but also person C.

We want to enable first order contact tracing (in the previous example: A's contacts: e.g. B), and additionally second order contact tracing (e.g. all of B's contacts, in our example: C). A user who receives a notification that there has been a contact with an infected person can then contact medical authorities and get the permission to upload their random TCNs to the server. Another possibility to proof contact with an infected person is using a so called zero-knowledge proof.

A zero-knowledge proof allows person B to proof that the server and B share a secret value without revealing any information aside the fact that they both know the value. In our case, B can prove knowledge of an infected TCN to the server without revealing the infected TCN. This can then authenticate B to upload their TCNs to the server.
 These TCNs can be marked separately, such that the user can get different notifications, depending if there was a direct confirmed exposure or an indirect exposure. 
 
 Tracing second order contacts increases significantly the number of traced potentially infected people. If every direct and indirect contact stayed in quarantine, a huge percentage of the population would be affected. Thus, tracing indirect contacts requires rapid testing of potentially infected people. For example, if person B in the above mentioned example is tested negative, person C does not require to stay in quarantine as the likelihood that person C has been infected by person B is small. Based on the rapid test results, health authorities can then decide who needs to isolate and who can continue with everyday life as usual. This avoids quarantining large fraction of the population during advanced stages of an epidemic. 

\section{Attack scenarios}\label{sec:attacks}
We evaluate possible attack scenarios in this section. These attack scenarios are not theoretical. Reports form South Korea \cite{SouthCoreaNature2020}\cite{SouthCoreaGuardian2020} show that attacks are done against users and that a state adversary is using the data to invade users' privacy. This emphasizes the importance of an approach, which collects minimal data and where such attacks are prevented by design. More attack scenarios will follow in a second draft of this document. 

\subsection{Reveal identity of infected person}\label{sec:id_infected_person}
A possible attacker could get the identity of infected people if they can directly access the TCNs of the infected people on the server (see \cref{sec:direct_download}). The attacker can install the app on a device and install the device somewhere together with a video camera. The camera records the people passing by and the device records the peoples' advertised TCNs together with the time. After some days, one of the people who passed the device and the camera finds out that they have been infected and uploads their TCNs to the server. The attacker can now download them and compare them with the TCNs on their device. The app detects a match. Now the attacker can check the matching TCNs and access the time on the database. Then, the attacker checks the video feed at that time and can possibly identify the infected individuals. 

\bigbreak
\noindent
\textit{Defense:}
The use of private set intersection cardinality protects further the identity of infected people. The attacker can still query the server multiple times to find out which of their contact TCNs was infected. Thus, a rate of queries needs to be limited to make these attacks more expensive. This type of linkage attack is in general possible with any proximity tracing app, which exchanges TCNs and notifies users. Any tech-savvy user can either use several devices, register their app mulitple times, modify the app and record the identities of other users. With PSI-cardinality and a rate limit, this attack becomes difficult and expensive. 

\subsection{Rebroadcasting of TCNs}
Attackers can use received contact TCNs and rebroadcast them as their own. Thus, people could get notifications of exposure triggered by falsely broadcasted TCNs even though they have not been at risk. In case of an infected attacker, the users would not get any notification, because the wrong TCNs, not the attackers' own TCNs, have been advertised.

\bigbreak
\noindent
\textit{Defense:}
Each TCN is concatenated with the time interval number. This time interval number is valid for ten minutes. If an attacker received a TCN from a neighbouring phone, they could rebroadcast this TCN to other users. In case this TCN is later marked as infected, only users who received this TCN from the attacker within the ten minute time interval will get a notification. The introduction of the time interval number reduces the validity of each TCN and thus puts a limit on this rebroadcasting attack.  
In the case of an infected attacker, the users would indeed never get notified of their possible exposure, since the attacker uploads their own TCNs. This would be the same if a person had their Bluetooth switched off, refused to upload their TCNs, or if they had not installed the app at all. 

\subsection{Upload of somebody else's TCNs}
An attacker could try to upload somebody else's TCNs in order to mark them as infectious. 
\bigbreak
\noindent
\textit{Defense:}
When receiving TCNs, the key needed for TCN generation is not transmitted. Thus, if the user is asked to upload their pseudo random TCNs, they need to provide the key they used for generating these TCNs. The server or the medical personels directly generates the TCNs with the keys, such that the TCNs are verified and the attacker cannot simply exchange TCNs. The attacker could of course try to provide fake keys, in which case the uploaded TCNs would not lead to any encounter notification, because the TCNs were never broadcasted.

\subsection{Forcing someone else to quarantine by self-reporting}
If an attacker wants to force someone to quarantine (e.g. manipulating sport events or annoying a neighbour), they can get into close proximity to this person and then try to self-report themselves as infected. 
\bigbreak
\noindent
\textit{Defense:}
The server is not directly accessible for the user. A user can only connect to the server if they have been granted access by medical personnel (by e.g. a TAN or a QR code). Medical personnel only grant access if the user has tested positive for COVID-19. The only case where self reporting is possible, is if the user has been exposed to an infected user, got the notification and can prove to the server with a zero-knowledge proof that they know an infected TCN. This knowledge serves as authorization to the server and allows the user to upload their own keys for TCN generation. These can be marked as second order TCNs and enable second order tracing of contacts as described in \cref{sec:2ndorder}.

\section{Related work and comparison}\label{sec:related_work}
 We present a summary of current suggested methods and protocols in this section. We also highlight major differences to our approach and where we improve privacy and security aspects. All approaches, except the last one in \cref{sec:pepppt}, are similar to our desing of DCTS. A broader overview of current efforts (including location based approaches) can be found in \cite{TCNresearchdoc}.
 
 \subsection{TCN coalition}\label{sec:tcn}
 We present the TCN protocol \cite{TCNProtocol} of the TCN Coalition, which we are also part of, in this section. Smartphones generate periodically changing TCNs and advertise them via Bluetooth. Neighboring devices store observed TCNs. An infected user uploads the generated TCNs to a server together with an additional memo field where report data can be written. All users then download the list of reported TCNs and check whether they have been exposed. This proposal provides server privacy and receiver privacy, however it does not provide source integrity and is vulnerable to linkage attacks.
 
 The TCN generation uses two report keys, a report authorization key $rak$, and a report verification key $rvk$, which are used to compute an initial temporary contact key (TCK):
 \begin{equation}
     tck_0 \leftarrow \mathrm{H}_{\mathrm{tck}}(rak), 
    \end{equation}
    \begin{equation}
    tck_i \leftarrow \mathrm{H}_{\mathrm{tck}}(rvk ~||~ tck{i-1}), 
 \end{equation}
 with H$_{\mathrm{tck}}$ as a domain-separated hash function with an output of 256 bit, and $i \geq 1$. The TCN is then generated with
 \begin{equation}
     tcn_i \leftarrow \text{H}_{\text{tcn}} (\text{le}_{\text{u16}}(i) ~||~ tck_i),
 \end{equation}
 with $H_{\mathrm{tcn}}$ as a domain-separated hash function with and output of 128 bit. In this scenario, everyone who knows $rvk$ and $tck_i$ can generate all subsequent $tcn_j$.
 
 In case of infection a user can generate a report for the period of $j1$ to $j2$:
 \begin{equation}
     report \leftarrow rvk ~||~ tck_{j1-1} ~||~ \text{le}_{\text{u16}}(j1) ~||~ \text{le}_{\text{u16}}(j2) ~||~ memo,
 \end{equation}
 with $memo$ as byte string (2-257 bytes). The memo field can contain any messages, for example self-reported symptoms. Then the user produces a signature $sig$ (using $rak$) for the $report$ and provides $report~ ||~ sig$ to the backend. Users can then verify source integrity by checking the signature $sig$ over $report$ using the patient's $rvk$.
 
The $rak$ needs to be changed frequently in order to protect better against linkage attacks; a maximal report time span of 6 hours or less is suggested. In this approach, each user can access the clear text of the patients' TCNs. This makes this design vulnerable against linkage attacks as described in \cref{sec:attacks} and also against replay attacks, where an attacker re-advertises someone else's TCNs. In our approach, private set intersection cardinality (as suggested in \cref{sec:private_set_intersection}) complicates linkage attacks even further and provides additional protection of the patient's identity and privacy.

 \subsection{DP-3T}\label{sec:dp3t}
 Another decentralized approach is presented in \cite{DP3Twhitepaper} from the DP-3T group. DP-3T presents two approaches, one with lower-cost and another one with increased privacy. We summarize both approaches in the following.
 
 \begin{itemize}
 
 \item \textbf{Low-cost decentralized proximity tracing:}
 Smartphones locally generate ephemeral identifiers $EphIDs$ (corresponds to TCNs in our approach), change them frequently and broadcast them via Bluetooth. Neighbouring smartphones store the observed $EphIDs$ together with duration and coarse time indication. A diagnosed patient gets authorization by health authorities and uploads a representation of their $EphIDs$. User query the server and get the patients' $EphIDs$. The smartphone then computes the risk score and in case notifies the user.
 
 The secret key $SK_t$ used for $EphID$ generation is rotated every day with
 \begin{equation}
     SK_t = H(SK_{t-1}),
 \end{equation}
 with $H$ as a cryptographic hash function. The $EphIDs$ are generated with $SK_t$ at the beginning of each day $t$:
\begin{equation}
    EphID_1 = || ... || EphID_n = PRG( PRF(SK_t, broadcast Key) ),
\end{equation}
using a pseudo-random function $PRF$ (e.g. HMAC-SHA256), a fixed public string $broadcast Key$, and a stream cipher $PRG$. Each $EphID$ is then broadcasted for one minute.

When infected, the backend collects the patients' $SK_t$ and $t$ and provides the data to users. Each smartphone then reconstructs the $EphIDs$ of infected users and checks whether the user has been exposed. This check is limited to a single day, in oder to increase efficiency for lookups and also to limit relay attacks (attacker redistributes captured $EphIds$). The smartphone then determines the user's risk score and notifies the user in case the score exceeds a threshold. 

This approach also offers various additional functions, e.g., storing the country a user has visited to ensure interoperability between countries. Also, the user can opt-in to share data with epidemiologists to support research. Location data or precise timing information will not be shared. 
The fact that the user receives a list of all the patients' $EphIDs$ makes this approach vulnerable e.g., against linkage attacks, where patient's $EphIDs$ could be linked to the patient's identity. 

     \item \textbf{Unlinkable decentralized proximity tracing:}
 A second design provides better privacy properties, however it requires the users to download larger volumes of data. The patients' $EphIDs$ are not revealed to the users, instead the $EphIDs$ on the backend are hashed and stored in a Cuckoo filter. This also allows infected users to not upload their data for sensitive locations or times. The general approach remains the same. Smartphones generate and broadcast $EphIDs$. The $EphIDs$ are generated for each epoch broadcasted time period $i$:
 \begin{equation}
     EphID_i = \text{TRUNCATE128}( H ( seed_i )),
 \end{equation}
 with $H$ as a cryptographic hash function truncated to 128 bit (TRUNCATE128). 
 
 Neighbouring smartphones observe the $EphID$ and store it as $H (EphID || i )$, with $H$ as a crpytographic hash function. The proximity, duration of encounter and a coarse time indication (e.g. day) is stored as well. 
 
 When diagnosed, patients upload ${(i, seed_i)}$ and can also choose for which epochs they want to reveal their $EphIDs$. The backend then computes
 \begin{equation}
     H ( \text{TRUNCATE128} (H (seed_i )) || i )
 \end{equation}
 and inserts the result into a Cuckoo filter. The filter is then sent to all users. The users' smartphones then applies the Cuckoo filter to their stored observed $H (EphIDs || i )$ and can then determine whether the user has been in contact. The risk score and notification remain the same as in the previous low-cost approach. 
 
 This design also offers the possibility to opt-in to share data with epidemiologists to support research. Location data or precise timing information will not be shared. 
 
 This design offers better protection of the infected users' identities. However, it requires the download of more data compared to the first low cost approach. 
 Still, the user has the cuckoo filter of all the patients' $EphIDs$ locally on their phone. A tech-savvy attacker can also determine which specific entries of their observed $EphIDs$ belong to infected users, e.g. by applying the Cuckoo filter to each entry individually and checking for overlaps. Our approach using private set intersection cardinality presented in \cref{sec:private_set_intersection} provides additional protection of the users' identities. 

 \end{itemize}
 
 An approach to protect from short-term and remote eavesdropping is also presented in \cite{DP3Twhitepaper}. They propose the use of Secret Sharing, where each $EphID$ is spread across $n$ beacons. Another user needs to receive at least $k$ shares in order to properly reconstruct the advertised $EphID$. Thus an attacker would need to be close to the user for a certain period of time in order to receive $k$ shares of the $EphID$.
 
 \subsection{Apple \& Google}\label{sec:appleandgoogle}
 Apple 's and Google's protocol is presented for example in \cite{appleGoogleTracing2020}\cite{appleAPI2020}\cite{googlecrypto2020}\cite{googleAPI2020}. Each phone generates a Temporary Exposure Key (TEK) valid for 24 hours. The key generation uses a cryptographic random number generator. With the TEK the phone generates a Rolling Proximity Identifier Key (RPIK). The Rolling Proximity Identifiers (RPI) are then derived with the RPIK. For this, each device calculates an interval number (ENIN):
 \begin{equation}
     ENIN(timestamp) \leftarrow \dfrac{timestamp}{60 \cdot 10},
 \end{equation}
 which provides a number for each ten minute time window. In this protocol, the interval number depends on when the key was first generated. Each key is valid for 24 hours, corresponding to 144 time intervals. We define $i$ as the number of ten minutes intervals since key generation:
 \begin{equation}
     i \leftarrow \dfrac{ENIN(time At Key Generation)}{144} \cdot 144.
 \end{equation}
 The RPI for a time $j$ where the identifier is calculated (Unix Epoch Time) is calculated with:
 \begin{equation}
     \text{RPI}_{i,j} \leftarrow AES_{128} (RPIK_i, \text{PaddedData}_j).
 \end{equation}
The PaddedData is a sequence of 16 bytes:
 \begin{itemize}
     \item PaddedData$_j$[0..5] = UTF8(\enquote{EN-RPI}) \\
     \item PaddedData$_j$[6..11] = 0x000000000000 \\
     \item PaddedData$_j$[12..15] = ENIN$_j$
 \end{itemize}
 
 They also offer the possibility to encrypt additional metadata along with the RPI. This metadata can then only be encrypted if the broadcasting has been infected and revealed their TEK. 
 
 A user that was infected and tests positive uploads their TEKs and the ENIN $i$ where the key validity started to a server. This upload can only be allowed by an official public health authority. The server distributes the keys and distributes them to the users. Each user then derives the infected person's RPI with the TEK and $i$. Afterwards, they match each of the infected RPIs with the encounteres identifiers. They allow for a two hour tolerance between when the RPI was supposed to be broadcasted and the actual scan time. If the exposure succeeds a threshold (based on exposure time and proximity), the user receives a notification.
 
 In the current design, users have no direct access to the encountered RPI and the infected RPI. Only a user with root access to the phone could possibly access this information and perform a linkage attack. Rebroadcasting other users' RPIs (if accessible with root access) would be possible within two hours time. This provides better protection than allowing attackers to directly access the patients RPIs and the encountered RPIs. Nonetheless, for effectively containing epidemic spread second order contact tracing (see \cref{sec:2ndorder}) needs to be introduced. Only then an infection chain can actually be interrupted.

 \subsection{Epione}\label{sec:coepi}
 In \cite{CoEpi} a decentralized approach is presented, which is also exchanging TCN via Bluetooth. In case of infection, the patient uploads their seed with which the advertised TCNs were computed to a server. There, the patient's TCNs are regenerated. They include private set intersection cardinality based on Diffie-Hellmann private set intersection as means to avoid linkage attacks, similar to our proposed solution in \cref{sec:private_set_intersection}. However, to actually stop the spread of SARS-CoV-2, second order tracing needs to be realized as well (see \cref{sec:2ndorder}).

 \subsection{ROBERT}\label{sec:pepppt}
ROBERT\cite{Pepp-robert} (ROBust and privacy-presERving proxmity Tracing) is the protocol suggested by Pan-European Privacy-Preserving Proximity Tracing (PEPP-PT). It is the only centralized protocol that we summarize in this paper. In this approach, the server keeps a record in a database for each registered app belonging to each user. For user $A$, this record comprises amongst other data a permanent identifier $ID_A$, which is assigned for each registered app only known to the server, a shared key $K_A$, and a list of exposed epochs. 

With ROBERT, the TCNs are generated by the server and sent to the app. The ephemeral Bluetooth identifier for A $EBID_{A,i}$ for epoch i are generated as:
\begin{equation}
    EBID_{A,i} = ENC(K_S, i | ID_A ),
\end{equation}
with $K_S$ as a server key stored by the server, and $ENC$ as a block-cipher with 64-bit block size. Additional to the $EBID$, each user also adds an encrypted country code $ECC$ and the time the message $M$ is broadcasted: $M_{A,i} = [ ECC_{A,i} | time_A]$. The advertised data $D$ consists of
\begin{equation}
    D_A = [M_{A,i} | \text{MAC}_{A,i}],
\end{equation}
with MAC as an HMAC-SHA256($K_{A, c_1} | M_{A,i}$) where $c_1$ is the prefix \enquote{01}.

Upon receiving $D_A$, another app $B$ retrieves $time_A$ from $D_A$ and obtains a timestamp $time_{A, B}$. The app then verifies that 
\begin{equation}
    | time_A - TRUNC_{16} (time_{A,B})| < \delta, 
\end{equation}
with $\delta$ as time tolerance (e.g. some seconds). If this is correct, the app stores $(D_A, time_{A, B})$ in its proximity list. 

If a user is infected, they can upload their proximity list for the time period where they have been infected to a server \cite{Pepp-app}, \cite{Pepp-robert}. The server verifies the uploaded data and checks whether a user is at risk to have been infected. It calculates a \enquote{risk score} depending on how long and how close a user has been with another infected person. User query status requests from the server regularly and get thus notified if their risk score exceeds a threshold. In this centralized approach, both server and user know whether they have had been in contact with an infected person. 

The tool is designed to work for different countries. If a server collects data with an country code from another country, it forwards the data to the respective server. 
 
 The server can link back the temporary identifiers to the permanent unique identifier linked to each user. The deanonymization of each user and also tracing users over time is thus trivial \cite{kuhn2020covid}. The users' contacts and social graphs are revealed to the server and an attacker with access to the server can exploit this sensitive information. Several other attacks are possible, such as linkability of contacts. A detailed security analysis can be found for example in \cite{DP3TRiskEvalPEPPPT}, and \cite{DP3TRiskEvalROBERT}. They conclude that this approach reveals many opportunities for exploitation and systematic misuse.



\section[Data protection assessment]{Data protection assessment\protect\footnote{Disclaimer: This assessment is preliminary and requires further elaboration and coordination on individual points.}}\label{sec:legal_aspects}

This section covers legal aspects, in particular the regulations in accordance with data protection law\footnote{It should be noted that apart from the strictly legal perspective, ethical and social implications will also have to be observed in the design of any digital contact tracing system.}. Insofar as the use of the DCTS app involves the processing of personal data, it must be compliant with the strict requirements of the GDPR.\footnote{The GDPR entered into force on 25 May 2018. It is directly binding and applicable to any entity that is processing the personal information of data subjects inside the European Economic Area (EEA), regardless of its statutory location and the data subjects' citizenship or residence.} This applies regardless of whether the app is operated by a public authority (at federal or state level) or a private institution. Essentially, the following questions arise:

\subsection{Applicability of GDPR}
The GDPR only applies to the processing of personal data, Art. 1 Sec. 1 GDPR. Conversely, the GDPR does not apply to purely factual or anonymous data. Art. 4 Sec. 1 GDPR defines personal data as \textit{\enquote{any information relating to an identified or identifiable natural person (‘data subject’); an identifiable natural person is one who can be identified, directly or indirectly, in particular by reference to an identifier such as a name, an identification number, location data, an online identifier or to one or more factors specific to the physical, physiological, genetic, mental, economic, cultural or social identity of that natural person}}.

It follows that the key threshold for determining whether data should be considered as personal data is not identification but rather identifiability of a specific natural person. Recital 26 GDPR stipulates that \textit{\enquote{[p]ersonal data which have undergone pseudonymisation, which could be attributed to a natural person by the use of additional information should be considered to be information on an identifiable natural person. […] The principles of data protection should therefore not apply to anonymous information, namely information which does not relate to an identified or identifiable natural person or to personal data rendered anonymous in such a manner that the data subject is not or no longer identifiable. This Regulation does not therefore concern the processing of such anonymous information, including for statistical or research purposes.}}.

Against this background, there is a strong indication that the GDPR would be applicable to the DCTS app. It cannot be ruled out that individual pieces of information in the processing chain will be personally identifiable by the use of additional information. This applies both to the advertised TCNs (which could in principle be attributed to a natural person by the use of the respective keys), the keys themselves, the -- implicit -- attribute \enquote{tested positive} as well as the determination of the \enquote{contact}. In addition, IP addresses
\footnote{As online identifiers, see \cite{GDPR_art4_lit1, GDPR_Recital_30}.}
as well as mac addresses\footnote{As device identifiers, see \cite{duesseldorfer_kreis}. } 
are typically considered as information on an identifiable natural person. This personal data will be processed, i.e. stored, uploaded and matched, at various stages during the operation of the DCTS app.
\footnote{In addition, the transmission of personal data within the framework of the Bluetooth connection may also occur, see \cite{zdnet}. } 

Conclusion: Personal data are processed and the GDPR is applicable.

\subsection{Lawfulness of data processing}
Pursuant to Article 6 GDPR, any processing of personal data requires an explicit authorisation. Art. 6 GDPR provides a whole range of possible justifications: from the consent of the data subject (i.e. the affected person) to a specific statutory regulation or a balancing of interests test. To the extent that health data are concerned (as would be the case with the -- implicit -- attribute \enquote{tested positive}), even stricter requirements must be met under Art. 9 GDPR. However, parallel with Art. 6 Sec. 1 Subsec. a. GDPR, Art. 9 Sec. 2 Subsec. a. GDPR provides a justification if the effective consent of all those affected is obtained.

As laid out above in Section 2, the DCTS app is conceived to function on a voluntary basis. This means that there should neither be a statutory obligation to use the app nor an automated implementation of the app on all end devices. In fact, the voluntary nature is a crucial factor for achieving widespread acceptance and trust among the population for any digital contact tracing system. 

It should be noted that some commentators have raised doubts whether a DCTS app can realistically be regarded as strictly voluntary because it may at least foster some form of indirect compulsion if the use of the DCTS app becomes a de-facto condition for taking part in public and social life.\cite{greenpeace} It is true that it cannot be ruled out that a private individual (e.g. a restaurateur) may practically \enquote{compel} its contractual partners to use the DCTS app. Therefore, it is all the more important that the non-use of the DCTS must not effectuate any weakening of an individual’s legal position vis-à-vis the state. Against this background, the European Data Protection Board\footnote{The European Data Protection Board (EDPB) is an independent European body, which contributes to the consistent application of data protection rules throughout the European Union, and promotes cooperation between the EU’s data protection authorities. The EDPB is composed of representatives of the national data protection authorities, and the European Data Protection Supervisor (EDPS). For further information see \cite{edpb}.} has already made clear that \textit{\enquote{[t]he use of such an application […] may not condition the access to any rights guaranteed by law.}}\cite{edpd_covid}

This means that, in principle, the DCTS app’s processing of personal data may be justified through formal consent of each and every individual app user pursuant to 6 Sec. 1 Subsec. a. GDPR, Art. 9 Sec. 2 Subsec. a. GDPR. However, it should be born in mind that the declaration consent must meet certain requirements. These are set out in Art. 7 GDPR (general conditions) and additionally in Art. 8 GDPR (particular conditions as regards minors) and Art. 12 GDPR (transparency conditions). These provisions in particular stipulate that
\begin{itemize}
    \item the declaration of consent must be explicit (\enquote{opt-in} in the context of the installation of the app on the smartphone); \\
    \item the declaration of consent must be free of any kind of compulsion (despite the urgency, no psychological pressure must be created, which leaves the individual hardly any real choice); \\
    \item sufficient, very comprehensible information is provided in an easy language that is appropriate for the addressee, based on which the addressee can form their decision (\enquote{informed consent}). This requirement is not at all trivial because, on the one hand, precise information must be provided about the purpose, means and all processing steps and circumstances, and, on the other hand, the addressee must not be overburdened.\\
\end{itemize}
    
Conclusion: If the requirements laid out above (and a few formalities not explicitly mentioned here\footnote{Cf. in detail Heckmann/Paschke, in\cite{GDPR_Ehmann}}) are complied with, the DCTS app’s processing of personal data may be justified under the GDPR.

\subsection{Further requirements under GDPR}
In addition, a whole range of procedural precautions must be observed in the process of developing the DCTS app. These include in particular
\begin{itemize}
    \item fulfilling certain information obligations (Art. 13 ff. GDPR), in particular with regard to the rights of data subjects, such as the right to information, the right to revoke consent, the right to deletion, the right to correction (e.g. when a \enquote{false-positive} test result is entered into the system); \\
    \item ensuring data protection through technology design and data protection-friendly default settings (Art. 25 GDPR); \\
    \item ensuring IT security (Art. 32 GDPR); \\
    \item providing proper information in case of data breach (Art. 33 GDPR); \\
    \item carry out a data protection impact assessment (Art. 35 GDPR); \\
    \item and several mores.\cite{ccc} \\
\end{itemize}{}

Conclusion: None of these requirements stand in the way of developing the DCTS app, but need to be incorporated from the outset of the design process to guarantee a compliant and thus sustainable use.

\subsection{Principle of proportionality}
In principle, the processing of personal data constitutes an encroachment on the fundamental right to informational self-determination. Even if this invasion can be justified (in this case: by consent, Art. 6 Sec. 1 Subsec. a. GDPR, Art. 9 Sec. 2 Subsec. a. GDPR), it is nonetheless subject to the principle of proportionality. This means that a legitimate purpose must be pursued through the use of the DCTS app and that the means used must be suitable, necessary and appropriate for achieving the purpose. The principle of proportionality must equally be observed in the case of \enquote{voluntary} (i.e. consented) app use, especially since the boundaries between voluntariness and compulsion may become blurred in the case of \enquote{urgent recommendations} on the part of the federal government as well as state governments and affiliated public institutions.

The purpose of this app is identifying chains of contacts to contribute to slowing down the spread of the virus (\enquote{flatten the curve}) in order to avoid overburdening the healthcare system while easing exit and contact restrictions. In this way, a balance between health protection and restrictions on fundamental rights should be achieved. At present, we still\footnote{Some EU member states have started a process of gradually easing some of the restrictions imposed. Nevertheless, the restrictions remain fairly unparalleled, both in terms of their range and comprehensiveness.} live under considerable restrictions on our fundamental rights (freedom of occupation, property, freedom of assembly, freedom of movement, freedom of religion, etc.).\footnote{In several recent decisions, the German Constitutional Court has made clear that under the current pandemic the scale will typically tilt in favor of the protection of public health when balancing the competing protected legal interests, see BVerfG, 7 April 2020, 1 BvR 755/20; BVerfG, 10 April 2020, 1 BvQ 28/20; slightly more in favor of the freedom of assembly BVerfG, 15 April 2020, 1 BvR 828/20.} The app is intended to help to reduce these restrictions, but at the same time provide sufficient health protection. This is undoubtedly a legitimate, and from a fundamental rights point of view, even welcome purpose.

The means used is a warning and protection system based on a disclosure of the fact of positively tested contacts, in order to give the persons affected the opportunity to take protective measures for themselves and third parties. In principle, this means is suitable to fulfil the purpose. Because of the voluntary nature of the system, it is currently unclear whether the DCTS app will ultimately be successful. From a constitutional point of view, the principle of proportionality is only violated if a proposed means is \enquote{utterly unsuitable} (according to the German Federal Constitutional Court in its settled case law). Since it cannot be ruled out that the DCTS app will fulfil its purpose, it currently meets the suitability threshold. Nevertheless, further evaluating the efficiency of digital contact tracing by refining the statistic models is a key component of the ContacTUM group’s DCTS research efforts, not least in order to further inform the dynamic proportionality assessment.

The next step is to assess whether the concrete setup of the DCTS app is necessary to achieve its purpose. This would not be the case if there were milder means that would be equally suitable to fulfill the purpose. Such a milder means is not immediately apparent:\footnote{There seems to be widespread consensus that processing location and/or movement data is not necessary to fulfill the purpose of digital contact tracing systems, i.e. identifying chains of contacts to contribute to slowing down the virus.} Continuing the lockdown would make the app unnecessary, but would prolong the considerable restrictions on other fundamental rights and is therefore not a preferable alternative. Relaxing the current contact and exit restrictions without installing a digital contact tracing system seems equally risky, because it is to be feared that people are not sufficiently sensitive to necessary self-restrictions.

However, in terms of necessity, there may be a crucial distinction between the two fundamental approaches to digital contact tracing, i.e. centralized and decentralized data reconciliation:
\begin{itemize}
    \item The decentralized approach (as chosen in our proposed DCTS app) could be viewed as a milder remedy compared to the centralized approach. This is because with central data reconciliation, all advertised TCNs would regularly be uploaded and stored together on the authentication server, at least for a short time. Thus, the risk of re-identification of affected persons may be greater. This central server would create a special attack vector, which raises questions of IT security. \\
    \item In the same vein, the decentralized approach seems more in line with the aforementioned principle of data minimization \cite{GDPR_art_5_sec_1} – which is, in essence, a materialization of the principle of proportionality. In the decentralized approach, only the positive cases are stored on the central server, while the actual data reconciliation takes place on the app on the end device. From the perspective of an app user, this aspect could increase the willingness to participate in a digital contact tracing system. \\
    \item Although there are currently no plans to combine the use of the DCTS app with compulsory protective measures, such as home quarantine or testing, some people may fear that the fact that they have tested positive could become known to third parties. In the decentralized approach, this fact would only be visible on the smartphone of the person concerned, i.e. it would be \enquote{in their hands}. From a necessity point of view, this could also speak in favour of the decentralized variant. \\
    \item Achieving widespread acceptance and trust among the population is the key factor for any digital contact tracing systems’s success. Only if the potential users trust in the app’s privacy architecture, they will actually use and follow the app in their daily lives. Against this background, it is crucial to obtain the endorsement of trustworthy institutions like the data protection agencies. In their letter to the European Commision of 14 April 2020, the EDPB has made clear that \textit{\enquote{[i]n any case, the EDPB wants to underline that the decentralised solution is more in line with the minimisation principle.}}\cite{edpb_guidance}\\
    \end{itemize}

Conclusion: Assuming that centralized and decentralized approaches to designing a digital contact tracing system are equally suitable to contribute to slowing down the spread of the virus, there are good arguments that a decentralized approach (as chosen in the present case with the DCTS app) is the preferable approach both from a proportionality and data minimization point of view.

\subsection{Supplementary consideration: Does the DCTS app require a statutory basis?}
On 3 May 2020, a group of data protection experts published a legislative proposal for a German \enquote{Law on the introduction and operation of an app-based tracing of infection risks with the SARS-CoV-2 (Corona) virus}\cite{netzpolitik_vorschlag} (the DCTS Law Proposal). The DCTS Law Proposal covers the requirements, framework and limits for the installation, use and operation of a DCTS app (Sec. 1 Subsec. 1). With regard to Germany, the authors argue that such a statutory basis is required for any digital contact tracing system.\cite{netzpolitik_warum}

\subsection{The main provisions of the DCTS Law Proposal}
The DCTS Law Proposal lays out that the purpose of any DCTS app is to foster contact tracing within the framework of the German Federal Infection Protection Act and to enable earlier infection warnings (Sec. 1 Subsec. 2). In addition, users shall be enabled to take additional measures to limit further spreading of the pandemic and thereby prevent an overburdening of the of the healthcare system capacities (Sec. 1 Subsec. 3).

Sec. 3 stipulates the voluntary nature of the app. The installation or use of the app may not be brought about either directly or indirectly by any form of compulsion. It is also forbidden to link any direct or indirect advantages or disadvantages to the use or non-use of the app. In the event of a notification of a potentially infectious encounter, there is no obligation to enter into (self-)quarantine, to undergo a medical test for COVID-19 or to notify third parties. In addition, users must be able to deactivate, terminate or delete the app at any time.

Pursuant to Sec. 4, the app as well as the server infrastructure necessary for its operation are provided and operated by the Robert Koch Institute.\footnote{The Robert Koch Institute (RKI) is the government’s central scientific institution in the field of biomedicine; see \url{https://www.rki.de/}. } Sec. 5 specifies the technical details for establishing potentially infectious encounters as well as for the data storing and processing. 

Sec. 7 stipulates that any data (TCNs, keys as well as any meta data) created, advertised or received by the DCTS app must not be used for any purposes other than those specified in the DCTS Law Proposal. Sec. 8 requires a data protection impact assessment (cf. Art. 35 GDPR), Sec. 9 requires that the source code of the app be published.

Sec. 10 prescribes that any user notified of a potentially infectious encounter the user shall be entitled to an immediate medical test for COVID-19. Sec. 11 prevents potential DCTS app operators from blocking the development of alternative DCTS apps by third parties. In case a medical test for COVID-19 is carried out, Sec. 12 provides the users with the additional option of transmitting their test results to the DCTS app operator for research purposes. The data transmitted may only be used by the operator to analyse and improve the accuracy of the matching process.
Finally, the DCTS Law Proposal shall cease to have effect when the German Bundestag finds that an \textit{\enquote{epidemic situation of national significance}}\cite{infection_protection} no longer exists, but no later than one year after its entry into force (Sec. 14 Subsec. 2).

\subsection{Preliminary assessment}
The DCTS Law Proposal enhances the evolving academic discussion around digital contact tracing systems. Whether such legislation should actually be passed remains to be discussed. Many of the provisions have a merely declaratory value and are somewhat characterized by a lack of trust in public institutions. This lack of trust may be attributable to the (perceived) political contention of the last weeks (centralized vs. decentralized approach, voluntary nature vs. compulsion, confusion surrounding the parallel release of a data donation app, etc.). However, it is questionable whether such contention is sufficient to require legislative action. As regards the DCTS Law Proposal, four dimensions should be distinguished:

\begin{itemize}

    \item Provisions such as Sec. 3 (voluntary nature) have a merely confirmatory and declaratory character. Even without such a provision, the processing of personal data would require a justification under the GDPR. Art. 6 Sec. 1 Subsec. a. GDPR, Art. 9 Sec. 2 Subsec. a. GDPR provides for such a justification if the effective consent of all those affected is obtained (see above under 7.2). Such consent must, in accordance with Art. 7 Subsec. 1 GDPR, not only be voluntary, but revocable at any time. 
The same goes for the provisions on deletion, purpose limitation and data protection impact assessment (Secs. 6-8), which can equally be derived from the GDPR. \\
    \item The specification of the technical procedure (Sec. 5) is redundant in the sense that it adopts a preliminary consensus reached in the scientific and political debate. Putting such a (temporary) consensus into law may make it difficult to adapt the technological standard to new findings. \\
    \item Provisions such as Sec. 4 (assigning the Robert Koch Institute as operator) have more than declaratory effect, as such organizational determination is typically not provided for by law. \\
    \item Secs. 9-11 clearly go beyond applicable law: Neither the obligation to publish the source code nor the right to an immediate medical test for COVID-19 nor the requirement to approve alternative apps result from previous legislation. From a legal policy perspective, it remains to be discussed whether these regulations are desirable.\\
    
\end{itemize}
However, one aspect of the DCTS Law Proposal indeed seems counterproductive: the proposed statute’s automatic expiry after one year (Sec. 14 Subsec. 2) will not prevent the app itself from being continuously used (for example on the basis of consent, see above). On the contrary: only the restrictive regulations, such as those on rigorous earmarking, would then no longer apply.

Conclusion: All in all, the DCTS Law Proposal is an important contribution to the academic discussion and could contribute to the overall acceptance of a digital contact tracing system. However, there is a risk that initiating a legislative process at this stage may further delay the introduction of the DCTS app. Alternatively, similar effects may be achieved by other public trust-building measures (such as the publication of \enquote{best practices} by the federal government): \enquote{law in action} instead of \enquote{law in books}. An overarching, general legal framework for similar applications aiming at nationwide risk prevention could then be created in a next step. Such an approach could also counteract the criticism that the DCTS Law Proposal constitutes an impermissible \enquote{single case} legislation. At any rate, parliamentary involvement in this complex matter is certainly to be welcomed.

\section{Conclusion}\label{sec:conclusion}
We present an improved decentralized, privacy preserving approach of a digital contact tracing service. We protect the users' identities and their personal data and focus on privacy and IT-security concerns. We incorporate legal principles and requirements, such as the GDPR or the Council of Europe’s Convention 108, into the very design of our solution.

In the decentralized approach, each infected user has the choice to contribute to fighting the pandemic and provide their advertised random IDs. This information concerns the user themselves and nothing can be inferred about other people. The worst case scenario of a potential security breach is the identification of infected users. 

In the centralized approach, each infected user can upload their observed IDs. This information is not only concerning the infected user who is revealing their social graph. This information concerns also every person the infected user has met in the past two/three weeks, as at least parts of their social graphs are revealed as well. The worst case scenario of a potential security breach now spans an entirely different magnitude. A variety of information can be obtained from a social graph with severe consequences for the individual. Social graphs of people can lead to identification of members belonging to groups of minorities (religious or otherwise) or even uncovering and endangering e.g., journalistic sources, whistle-blowers or political activists. They might reveal educational and social status, political circles and opinions, religious believes, and further sensitive information about people. 


The tracing of infectious contacts, digital or not, is an epidemiological tool that works effectively only if coupled with the ability to test potentially infected people quickly. Besides, the impact of tracking on epidemic containment depends on many factors that can also change over time. For this reason, the appropriate use of contact tracing must be permanently monitored and optimized through the use of proper epidemiological models. 

We improve existing concepts through the use of private set intersection which provides better privacy for the infected users. Additionally we show a way to trace second order contacts in a decentralized system while protecting the privacy of the user. This approach also prevents attackers from uploading contacts without prior contact to an infected person.

\bibliography{references}

\end{document}